%% file: iAF_Hall_2_incl_supp.tex
\documentclass[10pt,aps,prl,twocolumn,superscriptaddress,showpacs,preprintnumbers]{revtex4-1}

\usepackage{graphicx}
\usepackage{upgreek}
\usepackage{slashed}
\usepackage{amsmath}
\usepackage{amssymb}
\usepackage{latexsym}
\usepackage[utf8]{inputenc}
\usepackage{todonotes}
\usepackage[format=plain, justification=RaggedRight,   
singlelinecheck=false]{caption}
\usepackage[justification=centering]{subcaption}

\usepackage[english]{babel}
\usepackage[babel,kerning=true,spacing=true]{microtype}

\usepackage[papersize={8.5in,11in}]{geometry}
\usepackage[section]{placeins}

\usepackage[colorlinks=true]{hyperref}  
\hypersetup{
    unicode=false,          
    pdftoolbar=true,        
    pdfmenubar=true,        
    pdffitwindow=false,     
    pdfstartview={FitH},    
    pdftitle={Hall effect and Fermi surface reconstruction due to spiral 
antiferromagnetic order in underdoped cuprates},    
    pdfauthor={Andreas Eberlein, Walter Metzner, Subir Sachdev, Hiroyuki 
Yamase},     
    pdfsubject={Transport in cuprates},   
    pdfkeywords={Cuprates} {Hall effect} {Spiral antiferromagnetism}, 
    pdfnewwindow=true,      
    colorlinks=true,       
    linkcolor=magenta, 
    citecolor=blue,        
    filecolor=magenta,      
    urlcolor=blue           
}

\newcommand{\etal}{\textit{et~al.}}
\newcommand{\ie}{\textit{i.\,e.}}
\newcommand{\bs}{\boldsymbol}

\begin{document}

\title{Fermi surface reconstruction and drop of Hall number due to spiral 
antiferromagnetism in high-$T_c$ cuprates}

\author{Andreas Eberlein}
\affiliation{Department of Physics, Harvard University, Cambridge MA 02138, USA}
\author{Walter Metzner}
\affiliation{Max Planck Institute for Solid State Research,
 D-70569 Stuttgart, Germany}
\author{Subir Sachdev}
\affiliation{Department of Physics, Harvard University, Cambridge MA 02138, USA}
\affiliation{Perimeter Institute for Theoretical Physics, Waterloo, Ontario, 
Canada N2L 2Y5}
\author{Hiroyuki Yamase}
\affiliation{National Institute for Materials Science, Tsukuba 305-0047, Japan}

\date{\today}

\begin{abstract}
We show that a Fermi surface reconstruction due to spiral antiferromagnetic
order may explain the rapid change in the Hall number as recently observed near
optimal doping in cuprate superconductors [Badoux~\etal, Nature \textbf{531},
210 (2016)]. The single-particle spectral function in the spiral state exhibits 
hole pockets which look like Fermi arcs due to a strong momentum dependence of 
the spectral weight. Adding charge-density wave order further reduces the Fermi 
surface to a single electron pocket. We propose quantum oscillation measurements 
to distinguish between commensurate and spiral antiferromagnetic order. 
Similar results apply to certain metals in which topological order replaces 
antiferromagnetic order.
\end{abstract}

\maketitle

{\it Introduction.---}
Cuprate superconductors evolve from a Mott insulator to a correlated metal with 
increasing hole doping $p$. Long ago it was suggested that this evolution 
involves a quantum critical point (QCP) near optimal doping, and that the 
associated fluctuations are responsible for the high critical temperature for 
superconductivity \cite{Broun2008,Sachdev2010c,Scalapino2012}. The existence and 
nature of this QCP has not been clarified yet, because it is masked by 
superconductivity. Recently, the normal ground state became accessible by 
suppressing superconductivity with high magnetic fields. Near optimal doping in 
YBCO, Badoux~\etal\ reported a rapid change of the Hall number $n_H = (R_H 
e)^{-1}$ with doping~\cite{Badoux2016}. A similar behavior consistent
with a drastic drop of the charge carrier density upon lowering the doping was 
found shortly after in the Hall number and the resistivity of several cuprate 
materials~\cite{Laliberte2016-arXiv,Collignon2016-arXiv}. These results suggest 
that a QCP at optimal doping is associated with the reconstruction of a large 
Fermi surface enclosing a volume corresponding to a density $1+p$ of empty 
states (holes) at large doping, to small pockets with a volume corresponding to 
a hole-density $p$ in the underdoped regime. Moreover, these experiments 
indicate that the QCP for the closing of the 
pseudogap~\cite{Badoux2016,Badoux2016a} is distinct from that for the 
disappearance of charge order~\cite{Ramshaw2015}.

The observed transition in the charge carrier density could be associated with 
the termination of novel pseudogap metals without magnetic 
order~\cite{Yang2006,Kaul2007,YQi2010,Chatterjee2016-arXiv} or a QCP 
at which charge-density wave (CDW)~\cite{Caprara2016-arXiv} or Neel-type 
antiferromagnetic (AF)~\cite{Storey2016} order disappears. However, there is 
experimental evidence at least for YBCO that magnetic order in the ground state 
of the underdoped regime is \emph{incommensurate}~\cite{Haug2009,Haug2010}. 
From theoretical arguments, incommensurate AF has been shown to be favorable 
long ago for weakly doped Hubbard and $t$-$J$ 
models~\cite{Shraiman1989,Machida1989,Kato1990,Kane1990,Schulz1990,Chubukov1992,
Chubukov1995,Kotov2004,Sushkov2004,Raczkowski2006,Igoshev2010}. Recent 
renormalization group calculations suggest that incommensurate AF can coexist 
with superconductivity in a broad doping range~\cite{Yamase2016}. The energy 
gain from the magnetic order is tiny beyond the underdoped regime, but it 
becomes much more robust when superconductivity is suppressed. This raises the 
question whether the transition in the Hall number as seen in experiment could 
be caused by incommensurate antiferromagnetic order.

In this letter, we show that a quantum phase transition from a paramagnetic
metal to a spiral antiferromagnetic metal may indeed give rise to a crossover 
from $1+p$ to $p$ in the Hall number as seen in cuprates~\cite{Badoux2016}. 
Moreover, we find that the single-particle spectral function exhibits hole 
pockets with a strong spectral weight anisotropy reminiscent of Fermi arcs. 
Additional charge-density wave order can lead to a single electron pocket, with 
no additional Fermi surfaces, as observed \cite{yuxuan_wang}. To discriminate 
incommensurate spiral from commensurate antiferromagnetic order we propose a 
quantum oscillation experiment. We also note that certain topological Fermi 
liquids \cite{Nambu2016} have charge transport properties nearly identical to 
those of metals with magnetic order. And so our transport results apply also to 
such states.

{\it Spiral states.---}
In the following we describe spiral antiferromagnetic states using the 
mean-field Hamiltonian~\cite{Igoshev2010}
\begin{equation}
 H_\text{MF} = \sum_{\bs k}
 \begin{pmatrix} 
 c_{\bs k\uparrow}^\dagger, c_{\bs k + \bs Q \downarrow}^\dagger
 \end{pmatrix}
 \begin{pmatrix}
 \xi_{\bs k} & -A\\
 -A & \xi_{\bs k + \bs Q}
 \end{pmatrix}
 \begin{pmatrix}
 c_{\bs k\uparrow} \\ c_{\bs k + \bs Q \downarrow}
 \end{pmatrix},
\label{eq:H_MF}
\end{equation}
where $\xi_{\bs k} = -2t (\cos k_x + \cos k_y) - 4t' \cos k_x \cos k_y - \mu$ 
is the fermionic dispersion, $A$ the antiferromagnetic gap and $\bs Q = (\pi - 
2 \pi \eta, \pi)$ the ordering wave vector. We choose the hopping amplitude 
$t=1$ as our unit of energy in all numerical results. Diagonalization of 
$H_\text{MF}$ yields $H_\text{MF} = \sum_{\bs k, i=1,2} E_{\bs k, i} a_{\bs k 
i}^\dagger a_{\bs k i}$, where
\begin{equation}
 E_{\bs k, 1/2} = \frac{\xi_{\bs k} + \xi_{\bs k + \bs Q}}{2} \mp
 \sqrt{\frac{1}{4} (\xi_{\bs k} - \xi_{\bs k + \bs Q})^2 + A^2} \, .
\end{equation}
The quasi-particle operators $a_{\bs k i}$ are related to the bare fermion
operators by $c_{\bs k \uparrow} = \sum_j U_{\bs k, 1j} a_{\bs k j}$ and
$c_{\bs k + \bs Q \downarrow} = \sum_j U_{\bs k, 2j} a_{\bs k j}$, where
\begin{equation}
 U_{\bs k} =
 \begin{pmatrix}
 \frac{A}{\sqrt{A^2 + (\xi_{\bs k} - E_{1,\bs k})^2}} &
 \frac{A}{\sqrt{A^2 + (\xi_{\bs k} - E_{2,\bs k})^2}} \\
 - \frac{A}{\sqrt{A^2 + (\xi_{\bs k} - E_{2,\bs k})^2}} &
 \frac{A}{\sqrt{A^2 + (\xi_{\bs k} - E_{1,\bs k})^2}}
 \end{pmatrix}
\end{equation}
is the orthogonal transformation that diagonalizes $H_\text{MF}$. In a spiral 
antiferromagnetic state, the magnetic moments rotate in the $xy$ plane and 
their directions are modulated by the wave vector $\bs Q$ as $\bs m(\bs R_i) 
\sim \cos(\bs Q\cdot \bs R_i) \bs e_x + \sin(\bs Q\cdot \bs R_i) \bs e_y$, where 
$\bs R_i$ is a lattice vector.

We make the ansatz $A(p) = \alpha \, (p^* - p)\, \Theta(p^* - p)$ for the 
doping dependence of the gap, motivated by results for the on-site magnetization 
in spiral states in the $t$-$t'$-$J$ model~\cite{Sushkov2004}. A similar linear 
doping dependence of the gap in underdoped cuprates is also found in resonating 
valence bond mean-field theories for the $t$-$J$-model~\cite{Yang2006} or the 
pseudogap energy scale seen in experiments~\cite{Tallon2001}. For every doping 
$p$, the incommensurability $\eta$ is determined by minimizing the free energy 
at fixed $A$. More details on the doping dependence of $\eta$ can be found in 
the supplementary material~\cite{Suppl}.

\begin{figure}
	\centering
	\begin{subfigure}[t]{0.49\linewidth}
		\includegraphics[height=1.41in]{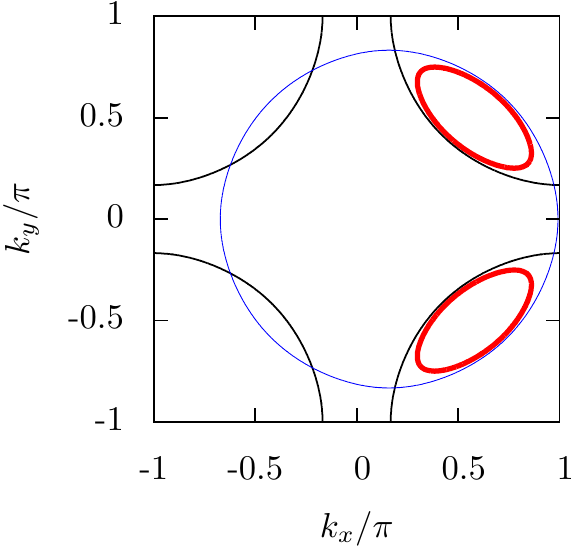}
	\end{subfigure}
	\begin{subfigure}[t]{0.49\linewidth}
		\centering
		\includegraphics[height=1.4in]{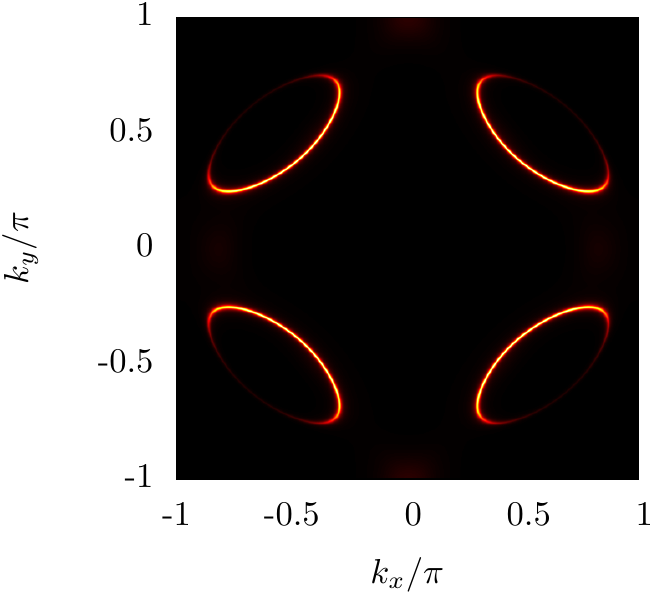}
	\end{subfigure}
	\begin{subfigure}[t]{0.49\linewidth}
		\includegraphics[height=1.41in]{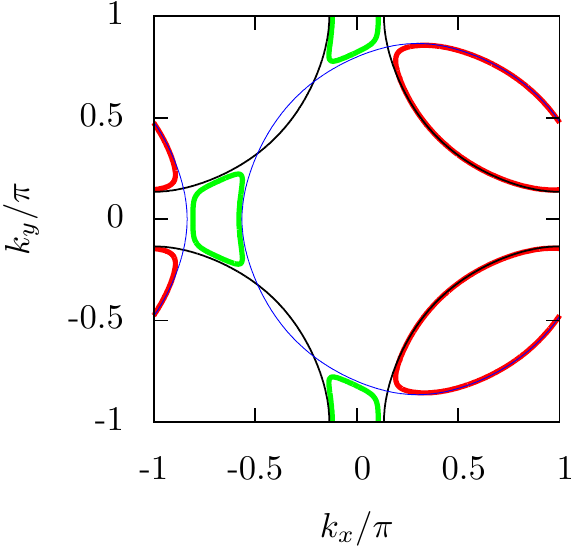}
	\end{subfigure}
	\begin{subfigure}[t]{0.49\linewidth}
		\centering
		\includegraphics[height=1.4in]{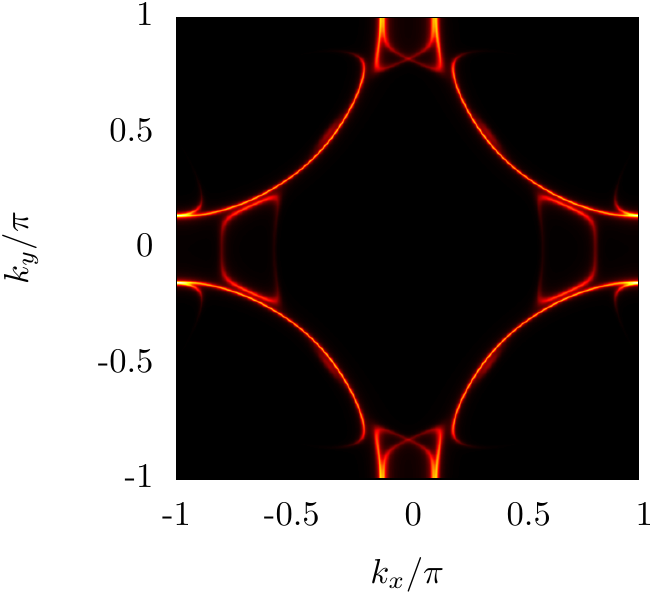}
	\end{subfigure}
\caption{Quasi-particle Fermi surfaces (left) and single-electron spectral 
functions (right) of spiral antiferromagnetic states for $p=0.08$, $A=0.63$ 
(top) and $p=0.15$, $A=0.23$ (bottom), where $t'=-0.35$ and $\eta \approx p$.
Hole and electron pockets in the left panels are marked in red and green, 
respectively, while the thin lines indicate the bare (black) and the $\bs Q$ 
shifted (blue) unreconstructed Fermi surfaces.} 
	\label{fig:A_FS}
\end{figure}

{\it Fermi surface and spectral function.---}
Filling the quasi-particle bands $E_{\bs k, 1/2}$ of the spiral state up
to the Fermi level yields hole and sometimes electron pockets as shown in
the left panel of Fig.~\ref{fig:A_FS}. For small doping one obtains only two 
hole pockets~\cite{Sushkov2004}, while for larger doping two electron pockets 
appear in addition. Spiral states with four hole pockets are also possible in 
principle \cite{Sebastian2008}, but were not obtained in the present study if 
the incommensurability $\eta$ is chosen such that the free energy is minimized.

The spectral function for single-particle excitations is given by $A(\bs 
k,\omega) = \sum_{\sigma} A_\sigma(\bs k,\omega)$, where
\begin{eqnarray}
 A_{\uparrow}(\bs k, \omega) &=&
 \sum_{i = 1,2} U_{\bs k, 1 i}^2 \,
 \delta(\omega - E_{\bs k, i}) , \\
 A_{\downarrow}(\bs k, \omega) &=&
 \sum_{i = 1,2} U_{\bs k - \bs Q, 2i}^2 \,
 \delta(\omega - E_{\bs k - \bs Q, i}) .
\end{eqnarray}
Numerical results for the spectral function at $\omega = 0$ are shown
in the right panel of Fig.~\ref{fig:A_FS} for two different hole dopings.
The momentum shift by $\bs Q$ in the quasi-particle bands contributing
to $A_{\downarrow}(\bs k,\omega)$ generates a shifted copy of all pockets.
The total (spin summed) spectral function is thus inversion symmetric,
but still exhibits a slight nematic deformation.

A most intriguing feature is that for small doping we obtain Fermi pockets
with a strongly suppressed spectral weight at their backside, reminiscent
of the mysterious Fermi arcs observed in underdoped cuprates. Let us see how 
this comes about for the hole-pockets related to particles with spin up. Their 
contribution to the spectral weight at $\omega=0$ is given by $U_{\bs k,11}^2 
\delta(E_{\bs k, 1})$, where $U_{\bs k,11}^2 = A^2/(A^2 + \xi_{\bs k}^2)$ for 
$E_{\bs k, 1} = 0$. From Fig.~\ref{fig:A_FS} it is clear that a large fraction 
of the inner side of the pockets is very close to the bare Fermi surface, where 
$\xi_{\bs k} = 0$. Hence $U_{\bs k,11}^2$ and thus the spectral weight there is 
almost one. The back side of the pocket is remote from the bare Fermi surface so 
that $A < \xi_{\bs k}$ and the spectral weight is thus quite small. A similar 
spectral function, albeit with fourfold rotation symmetry, is obtained in the 
commensurate case for $\eta = 0$.

{\it Hall coefficient.---}
The Hall coefficient is defined as $R_H = \sigma_H/(\sigma_{xx} \sigma_{yy})$,
where $\sigma_H$ is the Hall conductivity and $\sigma_{\alpha\alpha}$
is the longitudinal conductivity in direction $\alpha$. We compute the 
conductivities in a relaxation time approximation with a momentum independent 
scattering time $\tau$. Neglecting ``interband'' scattering between the two 
quasi-particle bands $E_{\bs k, 1}$ and $E_{\bs k, 2}$, the conductivities in 
the spiral state are given by the same expressions as for non-interacting 
two-band systems \cite{Voruganti1992}. Although the magnetic fields applied in 
the recent experiments by Badoux et al. \cite{Badoux2016} are impressively high, 
the product $\omega_c \tau$ is still small since the relaxation time $\tau$ is 
rather short ($\omega_c$ = cyclotron frequency). In the so-called weak-field 
limit $\omega_c \tau \ll 1$, one obtains \cite{Voruganti1992}
\begin{equation}
 \sigma_{\alpha\alpha} =
 e^2 \tau \sum_{i=1,2} \int \frac{d^2{\bs k}}{(2\pi)^2} \,
 \frac{\partial^2 E_{\bs k, i}}{\partial k_\alpha^2}
 n_F(E_{\bs k, i}) ,
\end{equation}
\begin{eqnarray}
 \sigma_{H} &=& -e^3 \tau^2 \sum_{i=1,2} \int \frac{d^2{\bs k}}{(2\pi)^2} \,
 \nonumber \\
 &\times& \! \Bigl[\frac{\partial^2 E_{\bs k, i}}{\partial k_x^2} 
 \frac{\partial^2 E_{\bs k, i}}{\partial k_y^2} - 
 \Bigl(\frac{\partial^2 E_{\bs k, i}}{\partial k_x \partial k_y}\Bigr)^2\Bigr]
 n_F(E_{\bs k, i}) .
\end{eqnarray}
Note that the $\tau$-dependence cancels in the Hall coefficient $R_H$.
The Hall number is defined as $n_H = (R_H e)^{-1}$. In special cases such
as parabolic dispersions, or for generic band structures with closed Fermi
surfaces in the high-field limit $\omega_c \tau \gg 1$, the Hall number is
simply given by the charge carrier density \cite{Ashcroft1976}.

\begin{figure}
	\centering
	\includegraphics[width=\linewidth]{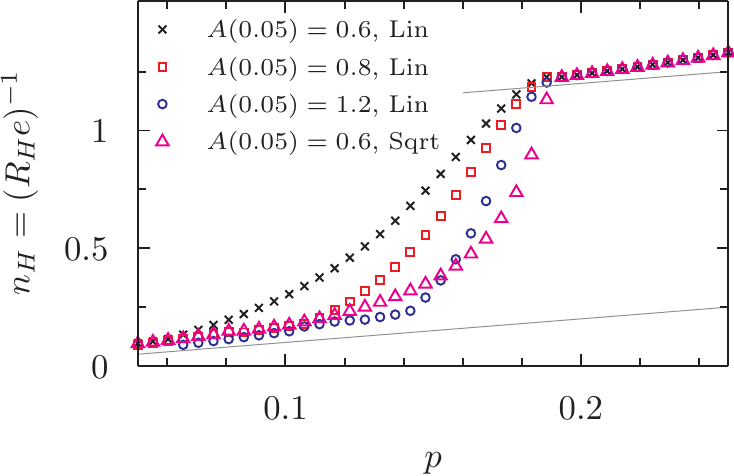}
	\caption{Hall number $n_H$ as a function of doping for $t' = -0.35$. Results 
for a linear dependence, $A(p) \sim (p^\ast - p)$, and a square root dependence, 
$A(p) \sim \sqrt{p^\ast - p}$, where $p^\ast = 0.19$ in both cases, are 
labeled as ``Lin'' and ``Sqrt'', respectively. The thin lines mark $n_H = p$ 
and $1+p$.}
	\label{fig:R_H}
\end{figure}
Results for the doping dependence of $n_H$ are shown for different values of the 
antiferromagnetic gap in Fig.~\ref{fig:R_H}. At small doping, $n_H$ is roughly 
given by the hole density $p$. Near $p^* = 0.19$, $n_H$ crosses over to $1 + p$. 
In the weak-field limit, the width of this crossover depends on the size of the 
antiferromagnetic gap. Larger gaps, or a square root doping dependence of the 
gap, lead to a sharper crossover. In the crossover region, the Fermi surface 
consists of hole and electron pockets, which is similar to the commensurate 
case and the YRZ scenarios studied in Ref.~\cite{Storey2016}.

In the high field limit $\omega_c \tau \gg 1$ and at zero temperature, $n_H$ is 
expected to be equal to the sum of the charge carrier densities of all Fermi 
pockets weighted by their sign, which is equal to the doping level $p$.
One would thus expect a jump in $n_H$ from $p$ to $1+p$ in the high-field
limit. The width of the crossover at weak and intermediate fields depends
on the Fermi surface geometry, temperature, and the field strength.

{\it Quantum oscillations.---}
\begin{figure}
	\centering
	\begin{subfigure}[t]{0.49\linewidth}
		\includegraphics[width=\linewidth]{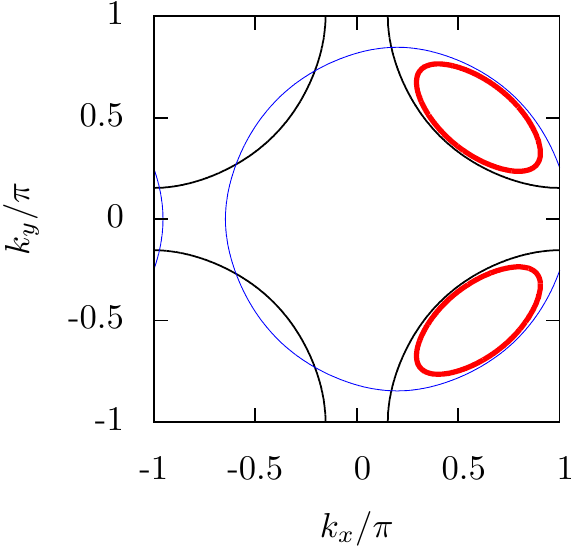}
		\caption{} \label{fig:FS_p0c10_iAF}
	\end{subfigure}
		\begin{subfigure}[t]{0.49\linewidth}
		\includegraphics[width=\linewidth]{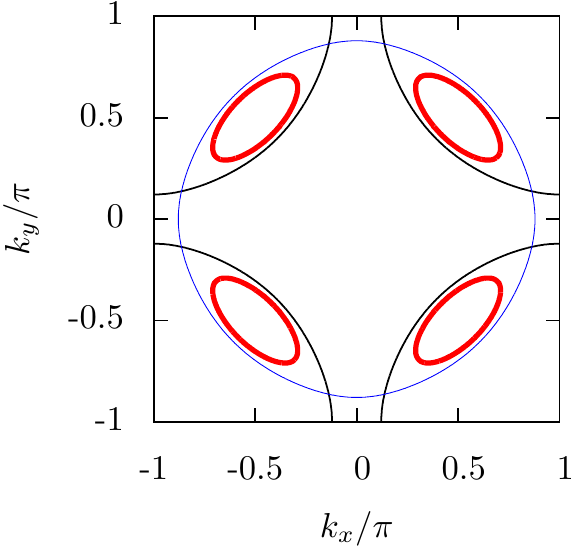}
		\caption{} \label{fig:FS_p0c10_cAF}
	\end{subfigure}
	\caption{Comparison between hole Fermi pockets of (a) incommensurate and (b) 
commensurate antiferromagnetic states with hole density $p = 0.1$ for $t' = 
-0.35$ and $A = 0.7$. $\eta = 0.1$ in a). }
	\label{fig:FS_Pocket_comparison}
\end{figure}
The measurements of the Hall coefficient by Badoux et al.~\cite{Badoux2016}
are consistent with both a commensurate N\'eel state and an incommensurate
spiral state. The Hall signal involves a sum over all Fermi surface sheets.
For sufficiently high fields, the Hall number is given by the sum over all
areas enclosed by the Fermi surface sheets, with electron-like Fermi surfaces 
counting negatively. Luttinger's theorem then implies that the Hall number is 
equal to doping $p$, irrespective of the incommensurability.

As an example, in Fig.~\ref{fig:FS_Pocket_comparison} we show Fermi surfaces
for a N\'eel state and an incommensurate spiral state at $p=0.1$ for parameters 
where only hole pockets appear. In the N\'eel state, the hole density is given 
by
\begin{equation}
 p = \sum_{\sigma = \uparrow,\downarrow}
 \int_{\text{MBZ}} \frac{d^2 \bs k}{(2\pi)^2} \, \Theta(E_{\bs k, 1}) =
 \int_{\text{BZ}} \frac{d^2 \bs k}{(2\pi)^2} \, \Theta(E_{\bs k, 1}),
\end{equation}
where integrals marked with MBZ and BZ are over the magnetic and full Brillouin 
zone, respectively. In the spiral state, one has
\begin{equation}
 p = \int_{\text{BZ}} \frac{d^2 \bs k}{(2\pi)^2} \, \Theta(E_{\bs k, 1}).
\end{equation}
The integrals measure the area of the hole pockets. The total area
is the same in both cases and is determined by $p$.

However, the single pockets in the spiral state are twice as large as the 
pockets in the N\'eel state. Spiral states could therefore be distinguished by 
quantum oscillations in the magnetic field dependence, as pointed out previously 
by Sebastian et al.~\cite{Sebastian2008}. For $\omega_c \tau > 1$, the magnetic 
susceptibility and other response quantities exhibit periodic oscillations as a 
function of $B^{-1}$ due to Landau quantization \cite{Ashcroft1976}. Each closed 
Fermi surface sheet yields a signal with an oscillation frequency
\begin{equation}
 F = \bigl(\Delta B^{-1}\bigr)^{-1} = \frac{\hbar S}{2 \pi e} ,
\end{equation}
where $S$ is the enclosed momentum space area. The pocket areas in the 
commensurate N\'eel state with four hole pockets and the incommensurate spiral 
state with two hole pockets are
\begin{align}
 S_\text{cAF} &= \Bigl(\frac{2\pi}{a}\Bigr)^2 \frac{p}{4} &
 S_\text{iAF} = \Bigl(\frac{2\pi}{a}\Bigr)^2 \frac{p}{2} ,
\end{align}
respectively, where $a$ is the lattice constant. The quantum oscillation 
frequencies of incommensurate spiral states are thus expected to be twice as 
large as those of N\'eel states at the same hole density. In particular, with 
the in-plane lattice constant of YBCO, $a = 3.8 \mathrm{\AA}$, one obtains the 
oscillation frequencies $F_\text{cAF} = 7160 \mathrm{T} \cdot p$ and
$F_\text{iAF} = 14320 \mathrm{T} \cdot p$.

\begin{figure}
	\centering
	\begin{subfigure}[t]{0.49\linewidth}
		\includegraphics[width=\linewidth]{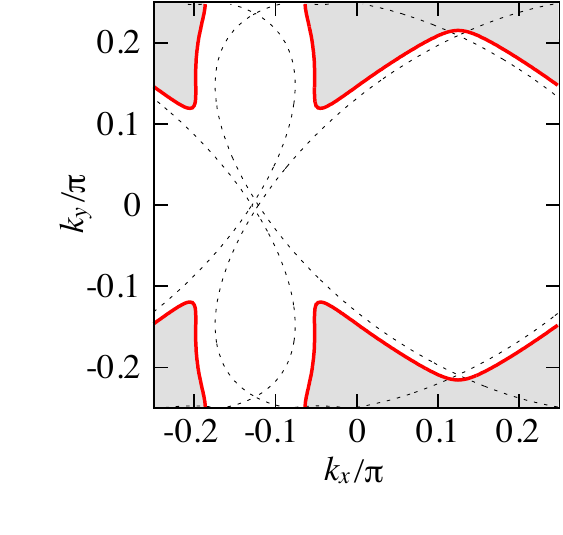}
		\caption{} \label{fig:A_iAF_sCDW}
	\end{subfigure}
		\begin{subfigure}[t]{0.49\linewidth}
		\includegraphics[width=\linewidth]{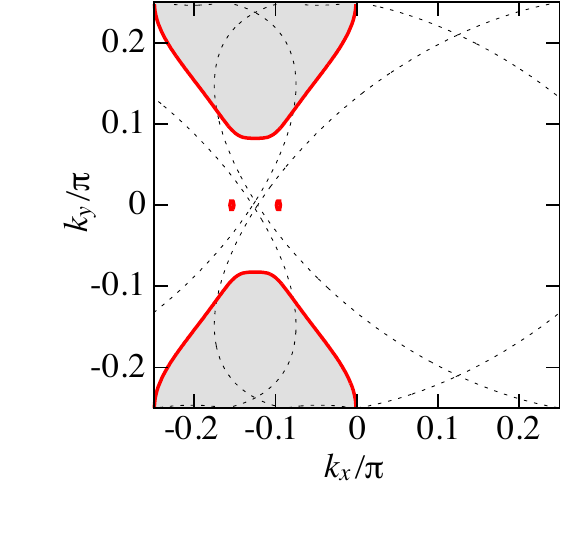}
		\caption{} \label{fig:A_iAF_dBDW}
	\end{subfigure}
\caption{Reconstructed quasi-particle Fermi surface due to spiral 
antiferromagnetic and charge-density wave order (red). The symmetry of the 
order parameter of the latter is \subref{fig:A_iAF_sCDW}) $s$-wave and 
\subref{fig:A_iAF_dBDW}) $d$-wave. The parameters are $t' = -0.35$, $p \approx 
0.12$, $\eta = 0.125$ and $A = 0.9$ for both and \subref{fig:A_iAF_sCDW}) $C = 
0.15$ and \subref{fig:A_iAF_dBDW}) $C = 0.72$. The black dashed lines 
show the Fermi surface of the spiral antiferromagnetic state after backfolding 
to the reduced Brillouin zone (\ie\ $C = 0$). The electron pockets are shaded 
in grey. }
	\label{fig:A_iAF_CDW}
\end{figure}
{\it Fermi surface for coexisting spiral antiferromagnetic and charge-density 
wave orders.---}
Cuprates show strong charge-density wave (CDW) correlations for $p \approx 
0.12$, which become long-ranged in high magnetic 
fields~\cite{TWu2011,TWu2013,Ghiringhelli2012,LeBoeuf2013,
Blanco-Canosa2013,Comin2015,Gerber2015}. In the field-induced 
ordered state, measurements of quantum oscillations and the Hall or Seebeck 
coefficient indicate a reconstruction of the Fermi surface into an electron 
Fermi pocket~\cite{Doiron-Leyraud2007,LeBoeuf2007,Doiron-Leyraud2013,Boebinger2016}, and 
no additional hole pockets are found in single-layer 
materials~\cite{Chan2016-arXiv}. Theoretical attempts to explain this 
reconstruction starting from a large hole Fermi surface~\cite{Allais2014} or the 
YRZ ansatz with small hole pockets~\cite{LZhang2016} yielded additional open 
Fermi surface sheets or hole pockets. A reconstruction into one electron pocket 
could work starting from four Fermi arcs~\cite{Harrison2016}, but that proposal 
did not answer the question about their origin.

Coexisting spiral AF and bidirectional CDW order can be described by adding
\begin{equation}
 H_\text{CDW} = - C \sum_{\bs k,\sigma,i} f\bigl(\bs k + \frac{\bs 
q_i}{2}\bigr) \bigl(c_{\bs k + \bs q_i\sigma}^\dagger c_{\bs k\sigma} + c_{\bs 
k\sigma}^\dagger c_{\bs k + \bs q_i\sigma}\bigr)
\end{equation}
to Eq.~\eqref{eq:H_MF}, where $C$ is the CDW order parameter. Bidirectional CDW 
order with ordering wave vectors $\bs q_1 = (\pi / 2, 0)$ and $\bs q_2 = (0, 
\pi/2)$ is chosen as a simple approximation for the (incommensurate) CDW 
with a period of roughly four lattice constants that is seen in experiments. 
The form factor $f(\bs k)$ is of predominantly $d$-wave symmetry ($f(\bs k) = 
\cos k_x - \cos k_y$) in cuprates. We determine the Fermi surface for this 
symmetry and an onsite CDW with s-wave symmetry ($f(\bs k) = 1$). In 
Fig.~\ref{fig:A_iAF_CDW} we show that CDW order of both symmetries can 
reconstruct the two hole Fermi pockets of the spiral state (similar to those 
in Fig.~\ref{fig:FS_p0c10_iAF}) into a single electron pocket. For 
$d$-wave CDW order with a smaller order parameter, the resulting Fermi surface 
is qualitatively similar to Fig.~\ref{fig:A_iAF_sCDW}~\cite{Suppl}. 
Intriguingly, larger $d$-wave CDW order parameters, as in 
Fig.~\ref{fig:A_iAF_dBDW}, can give rise to additional Dirac cones in the 
spectrum. These arise from the inversion of two bands with different spin 
chirality, similar to topological insulators with spin-orbit coupling.

{\it Conclusions.---}
We have shown that spiral antiferromagnetism may explain several 
features of the phenomenology of hole-doped cuprates. The spectral function of 
spiral antiferromagnetic states consists of hole pockets, which due 
to a strong momentum dependence of the spectral weight look like Fermi arcs. 
The Fermi surface reconstruction at a quantum critical point due to spiral 
antiferromagnetic order may explain the rapid change in the Hall number as 
recently observed near optimal doping in cuprate superconductors. 
In a doping regime where it is observed in cuprates, additional charge-density 
wave order further reconstructs the hole Fermi surface of the spiral 
antiferromagnetic state into a single electron pocket.

Metals with topological order can have the same charge transport properties as 
metals with magnetic order~\cite{Nambu2016}, but their fermionic quasiparticles 
carry a pseudospin with no Zeeman coupling, and so can be distinguished in 
quantum oscillation or low $T$ photoemission.

The detection of spiral antiferromagnetic order, or quantum-fluctuating order in 
the topological metals, in hole-doped cuprates near optimal doping would 
significantly improve our understanding of the cuprate phase diagram. 
Incommensurate antiferromagnetism is expected from a theoretical point 
of view and is favorable over N\'{e}el-type antiferromagnetism. We 
propose quantum oscillation measurements to distinguish between N\'{e}el-type 
and spiral antiferromagnetic order. 

{\it Acknowledgments.---} 
We would like to thank O.~Sushkov, L.~Taillefer and R.~Zeyher for valuable 
discussions. AE acknowledges support from the German National Academy of 
Sciences Leopoldina through grant LPDS~2014-13. HY appreciates support by the 
Alexander von Humboldt Foundation and a Grant-in-Aid for Scientific Research 
from the Japan Society for the Promotion of Science. This research was supported 
by the NSF under Grant DMR-1360789 and MURI grant W911NF-14-1-0003 from ARO. 
Research at Perimeter Institute is supported by the Government of Canada through 
Industry Canada and by the Province of Ontario through the Ministry of Economic 
Development \& Innovation. SS also acknowledges support from Cenovus Energy at 
Perimeter Institute.


\input{iAF_Hall_2_incl_supp.bbl}

\pagebreak
\begin{widetext}

\appendix
\onecolumngrid
\appendix
\section{\large Supplementary information}

In this Supplementary Material we provide details on the dependence of 
the spiral antiferromagnetic gap $A$ and the incommensurability $\eta$ on the 
hole doping concentration $p$. We also show the quasi-particle Fermi surface 
for a state with spiral antiferromagnetism and charge-density wave order 
with d-wave symmetry for an additional set of parameters.

In Fig.~\ref{fig:A}, we show the doping dependence of the spiral 
antiferromagnetic gap $A$ that we assume in the main text. In order to study 
the influence of how the antiferromagnetic gap closes, we consider gaps that 
vanish as $A(p) \sim p^\ast - p$ or $\sim \sqrt{p^\ast - p}$, with $p^\ast = 
0.19$.
\begin{figure}[h]
	\centering
	\includegraphics[width=0.4\linewidth]{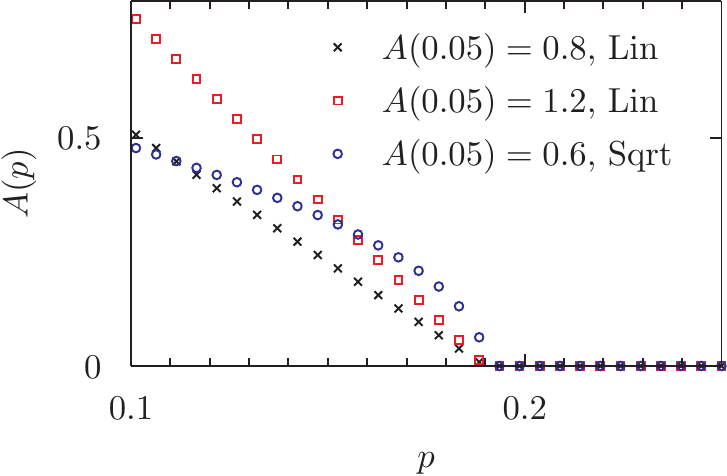}
	\caption{Antiferromagnetic gap $A$ as a function of doping $p$.}
	\label{fig:A}
\end{figure}

In Fig.~\ref{fig:eta}, we show the doping dependence of the incommensurability 
$\eta$ that parametrizes the antiferromagnetic ordering wave vector as $\bs Q = 
(\pi - 2 \pi \eta, \pi)$. $\eta$ was determined by minimization of the 
fermionic contribution to the ground state energy for any given $p$ and $A(p)$. 
For the parameters considered, $\eta$ is equal to $p$ to a very good 
approximation.
\begin{figure}[h]
	\centering
	\includegraphics[width=0.4\linewidth]{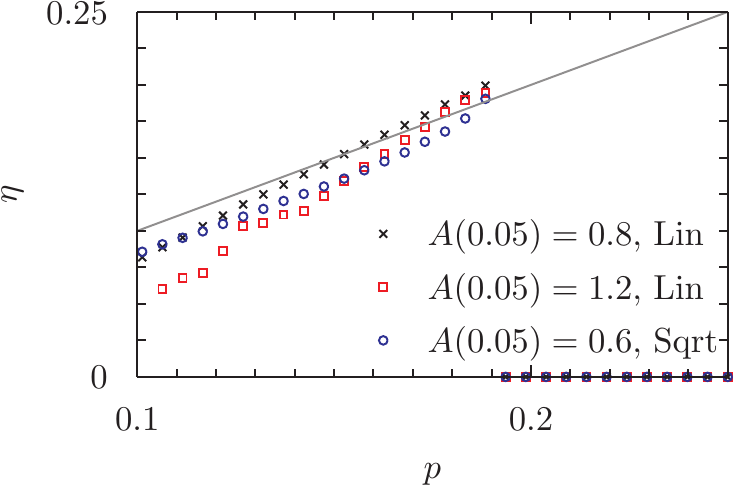}
	\caption{Incommensurability $\eta$ as a function of doping. The grey line 
marks $\eta = p$.}
	\label{fig:eta}
\end{figure}

In Fig.~\ref{fig:dCDW_noDirac}, we show the quasi-particle Fermi surface for 
coexisting spiral antiferromagnetic and charge-density wave order, the latter 
having $d$-wave symmetry. In comparison to Fig.~4b in the main text, the 
charge-density wave gap is smaller. The Fermi surface is reconstructed into one 
electron pocket, but no Dirac cones appear close to the Fermi surface.
\begin{figure}[h]
	\centering
\includegraphics[width=0.3\linewidth]{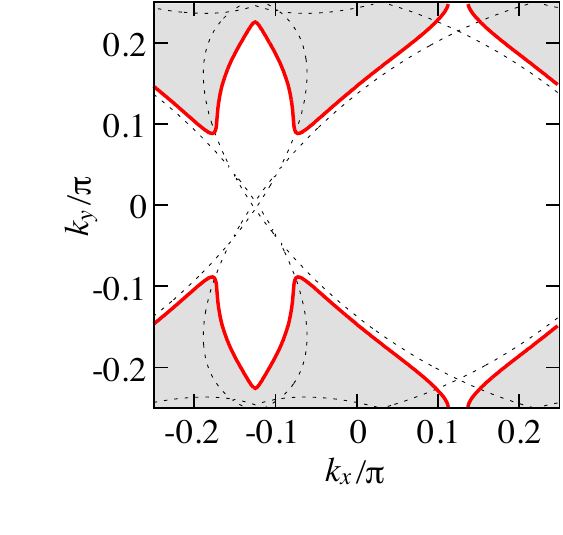}
	\caption{Reconstructed quasi-particle Fermi surface due to spiral 
antiferromagnetic and $d$-wave charge-density wave order (red). The parameters 
are $t' = -0.35$, $p \approx 0.11$, $\eta = 0.125$, $A = 0.8$ and $C = 0.3$. 
The black dashed lines show the Fermi surface of the spiral antiferromagnetic 
state after backfolding to the reduced Brillouin zone (\ie\ $C = 0$). The 
electron pocket is shaded in grey.}
	\label{fig:dCDW_noDirac}
\end{figure}

\vspace*{2cm}

\end{widetext}

\end{document}

%% file: iAF_Hall_2_incl_supp.bbl
%